\documentclass[doublecol]{epl3}
\usepackage{graphicx}
\usepackage{epstopdf}
\usepackage{bm}
\usepackage{subfigure}
\usepackage{sidecap}
\usepackage{graphicx}
\usepackage{epstopdf}
\usepackage{amssymb}
\usepackage{amsmath}
\usepackage{bm}
\usepackage{subfigure}
\usepackage{sidecap}
\graphicspath{{Figures/}}

\newcommand{\be}{\begin{equation}}
\newcommand{\ee}{\end{equation}}

\newcommand{\bea}{\begin{eqnarray}}
\newcommand{\eea}{\end{eqnarray}}
\newcommand{\bd}{\begin{displaymath}}
\newcommand{\ed}{\end{displaymath}}
\newcommand{\ba}{\begin{array}}
\newcommand{\ea}{\end{array}}
\newcommand{\bi}{\begin{itemize}}
\newcommand{\ei}{\end{itemize}}
\newcommand{\bc}{\begin{center}}
\newcommand{\ec}{\end{center}}
\newcommand{\bfl}{\begin{flushleft}}
\newcommand{\efl}{\end{flushleft}}
\newcommand{\bfr}{\begin{flushright}}
\newcommand{\efr}{\end{flushright}}

\newcommand{\bl}{\begin{aligned}}
\newcommand{\el}{\end{aligned}}

 \def\bd{{\bf d}}

\def\={\!\!\!&=&\!\!\!}
\def\+{\!\!\!&&\!\!\!+~}
\def\-{\!\!\!&&\!\!\!-~}

\title{Investigation of magnetic phases in parent compounds of Iron-chalcogenides via quasiparticle scattering interference}

\author{Bhaskar Kamble\inst{1}
\and Alireza Akbari\inst{1,2,3}
\and Ilya Eremin\inst{4,5}
}
\shortauthor{Kamble et al.}

\institute{
$^{1}$Asia Pacific Center for Theoretical Physics (APCTP), Pohang, Gyeongbuk, 790-784, Korea\\
$^{2}$Department of Physics, POSTECH, Pohang, Gyeongbuk 790-784, Korea\\
$^{3}$Max Planck POSTECH Center for Complex Phase Materials, POSTECH,
Pohang 790-784, Korea\\
$^4$Institut f\"{u}r Theoretische Physik III, Ruhr-Universit\"{a}t Bochum, 44801 Bochum, Germany\\
$^5$National University of Science and Technology ‘MISiS’, 119049 Moscow, Russian Federation
  }
\abstract{
We employ  a five-orbital tight-binding model to develop the mean field solution for various  possible spin density wave states  in the iron-chalcogenides. 
The quasiparticle interference (QPI)  technique is applied to detect  signatures of these   states due to scatterings arising from  non-magnetic impurities. 
Apart from the  experimentally observed   double striped structure with ordering vector $(\pi/2,\pi/2)$, the QPI method is investigated for the  extended-stripe  as well as the orthogonal  double stripe phase.
We discuss QPI as a possible tool to detect and classify various magnetic structures with different electronic structure reconstruction within framework  of the Fe$_{1+y}$Te compound.
}

\begin{document}
\maketitle
One of the biggest mysteries of the Fe-based superconductors is the striking difference between the magnetic phase of the pnictide and chalcogenide compounds. The parent compounds of the Fe-pnictide superconductors show a magnetic state with a $(\pi,0)$ spin density wave (SDW) vector which gives way to superconductivity upon doping and/or application of pressure~\cite{RevModPhys.83.1589}. The Fermi surfaces in both the pnictides and chalcogenides in general show common features such as hole pockets at the $\Gamma$ point and electron pockets at the X and Y points of the Brillouine Zone~\cite{PhysRevLett.100.237003,PhysRevLett.101.026403,PhysRevLett.101.057003,PhysRevLett.101.177005,PhysRevB.79.054517,0295-5075-83-4-47001,PhysRevLett.101.216402}. These pockets are well-nested at the wave vector of $(\pi,0)$ or $(0,\pi)$. This is also the wave vector of the SDW, and hence band nesting is commonly accepted as the mechanism of the SDW in the pnictide compounds~\cite{PhysRevB.81.024511}.

However the chalcogenides show a completely different magnetic structure, with a SDW wave vector of $(\pi/2,\pi/2)$ for zero or low doping, forming the so-called double-striped (DS) phase and hence the origin of the SDW in the chalcogenides cannot be explained by the nesting of the hole and electron bands~\cite{PhysRevB.79.054503}. The origin of the SDW in the chalcogenides is still under hot debate. Among the chalcogenides, the compound Fe$_{1+y}$Te poses a particular mystery. Apart from the fact that the SDW is different from the pnictides, it has been pointed out that a generic commensurate $(\pi/2,\pi/2)$ SDW is a superposition of ${\bf Q}_{1/2}=(\pi/2,\pm\pi/2)$ 
wave vectors~\cite{PhysRevLett.109.157206}. It was shown in the latter reference that quantum fluctuations in a localized model stabilize the so-called orthogonal  double stripe  (ODS) phase which contains both these wave vectors. 
Experimentally, it was shown that Fe$_{1+y}$Te tends to order in the ODS state based on the neutron scattering structure factor~\cite{PhysRevLett.107.216403}. Theoretical studies such as exact diagonalization~\cite{1742-6596-200-2-022058} and mean field studies of the $t$-$J$ model~\cite{PhysRevB.86.134512} also favor the ODS state. 
A recent spin-polarized scanning tunnelling microscopy (STM) study confirmed the DS  phase at low excess iron, but suggested the ODS phase at higher $y$ values~\cite{Enayat08082014}. The magnetic order becomes even more complex upon increasing the excess iron. 
In particular,  recent  neutron diffraction results show an  incommensurate SDW state~\cite{PhysRevLett.102.247001}, and also possibly a helical state~\cite{PhysRevLett.115.177203}.

\begin{figure}
\begin{center}
\includegraphics[scale=0.3,angle=0]{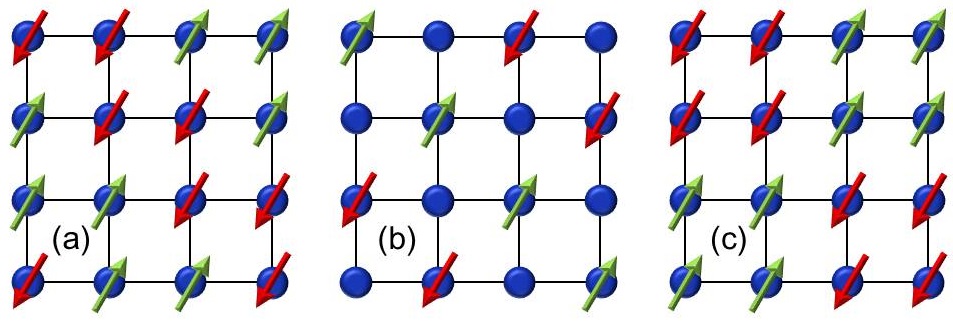}\\
\vspace*{-1.cm}
\hspace{0.45cm}
\includegraphics[scale=0.5,angle=0]{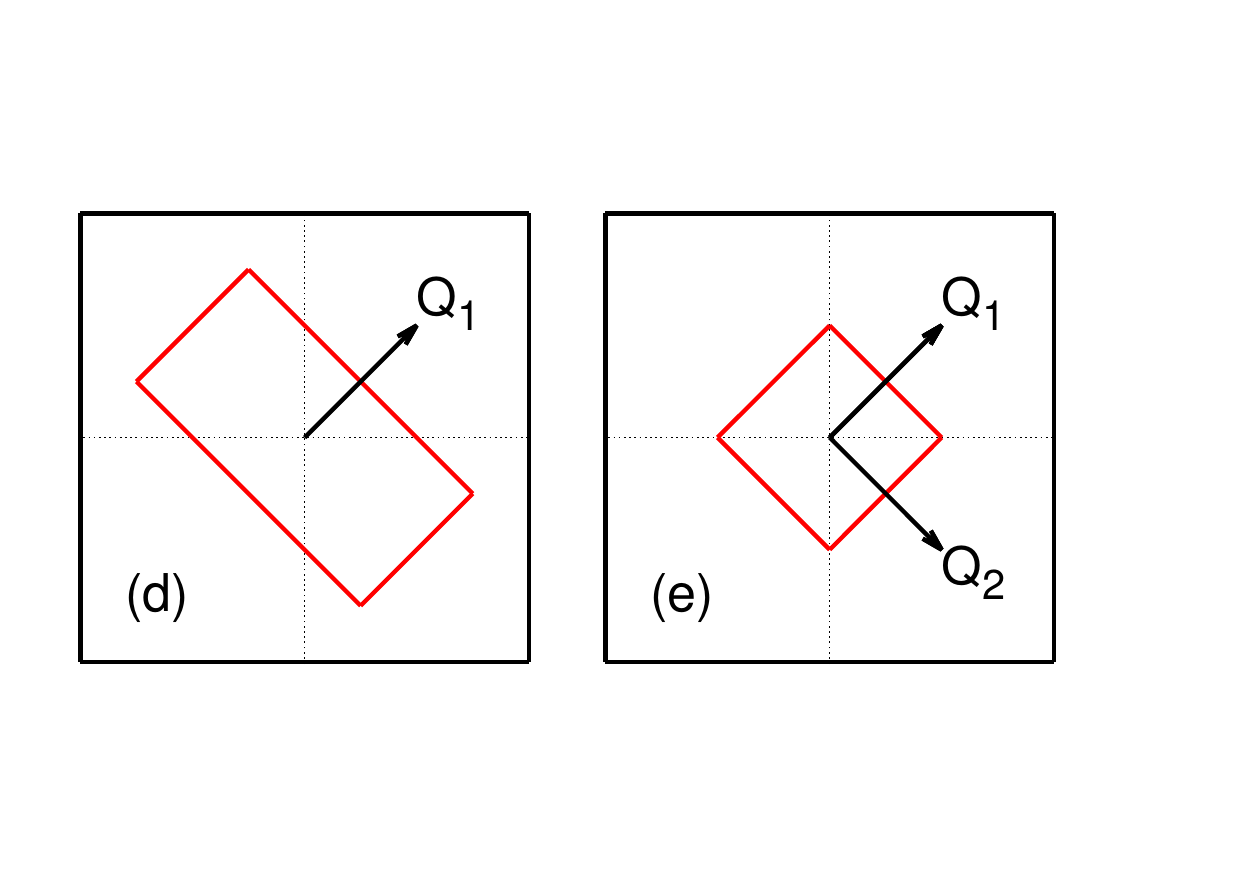}
\end{center}
\vspace{-1.6cm}
\caption{The magnetic structure with $(\pi/2,\pi/2)$ SDW wave vector with (a) double striped structure ($\phi=\pi/4$), and (b) extended-stripe phase ($\phi=0$). 
(c) Shows  the ODS phase with both $(\pi/2,\pi/2)$ and $(\pi/2,-\pi/2)$ ordering wave-vectors.
(d) The magnetic Brillouin zone (red rectangle) for an SDW with ${\bf Q}_1=(\pi/2,\pi/2)$ and (e) the magnetic Brillouin zone (red square) for an 
SDW with ${\bf Q}_{1/2}=(\pi/2,\pm\pi/2)$. The black square is the fulll Brillouin zone for the paramagnetic case.}
\label{fbzmbzc2c4}
\end{figure}

Due to the close proximity of magnetism and superconductivity in the iron-compounds, the superconductivity is believed to be mediated by spin fluctuations rather than lattice vibrations~\cite{0034-4885-74-12-124508}. Hence understanding the magnetic phase is important to understand the superconducting state. The investigations of the quasiparticle excitation can give valuable information on the nature of superconductivity since they are intimately connected to the superconducting gap, $\Delta_{\bf k}$. These quasiparticles can be detected by momentum-sensitive techniques such as ARPES to find the quasiparticle spectrum and the quasiparticle density of states. Among phase-sensitive techniques such as Josephson tunnelling, SQUID interferometry, etc., the one that directly investigates quasiparticle effects is the quasiparticle interference (QPI) spectroscopy based on scanning tunneling microscopy (for a review see~\cite{0034-4885-74-12-124513}). In this method, interference of quasiparticles due to random impurities in the sample is detected through the spatial modulation of the local quasiparticle density of states at a constant energy bias. This density of states corresponds to the differential conductance which is measured by STM, and the Fourier transform of these real space data then gives the wave vectors at which the dominant scatterings occur. QPI is a powerful tool because it simultaneously yields energy-dependent real-space and momentum-space information on the quasiparticle wave-functions, scattering processes and coherence factors. This information can be used to distinguish between different superconducting 
order parameters~\cite{0034-4885-74-12-124513,PhysRevLett.71.3363,0295-5075-85-3-37005,Hanaguri23042010,Hanaguri13022009,Allan:2012aa,Chi:2014aa,Akbari:2010aa,PhysRevLett.104.257001,Yamakawa:2015aa,Hirschfeld:2015aa,Sykora:2011aa,Akbari:2011aa,0295-5075-100-3-37004,Akbari:2013aa,0295-5075-103-2-27004,0295-5075-100-3-37002,Akbari:2014aa,doi:10.7566/JPSJ.83.061015}. In addition, QPI can also access the momentum space structure of the unoccupied states that are inaccessible to photoemission~\cite{McElroy1}.

Given the complicated scenario regarding the magnetic phases of Fe$_{1+y}$Te, it is natural to ask if one can apply a non-magnetic method to detect these phases. In this article we explore the possibility of using non-magnetic-QPI as a possible tool to detect the magnetic phases of this compound. QPI can be a valuable tool to characterize the different possible magnetic structures since the magnetic states will influence the band structure which will in turn influence the QPI features. To illustrate the possibility, we undertake a theoretical investigation of QPI for three different kinds of magnetic structures involving the  ${\bf Q}_1$ and ${\bf Q}_2$ wave vectors. For low $y$, only one of either ${\bf Q}_1$ or ${\bf Q}_2$ is present with a phase-angle  $\phi=\pi/4$ which results in the DS phase shown in Fig.~\ref{fbzmbzc2c4}(a). This is the first phase that we consider. For the same wave vector, one can also theoretically change the magnetic structure by varying the phase angle $\phi$, which thus offers a continuous parameter which determines the magnetic structure. The second magnetic phase we consider is with ${\bf Q}_1$ and $\phi=0$. The third magnetic phase we consider is the ODS phase discussed in~\cite{PhysRevLett.109.157206}. We consider the five-orbital tight-binding model of Ducatman, et. al\cite{PhysRevB.90.165123}. This model is meant for the Fe$_{1+y}$Te compounds and explicitly considers the effect of the additional interstitial Fe atoms. \\

{\it  Spin density wave with  ${\bf Q}_1$ wave vector:}
\label{secone}
 In the following, we present  the mean field frame work  for general SDW with ${\bf Q}_1$.
For this, the mean-field ansatz for the electronic density in orbital $\gamma$ and site $i$ can be written as 
%
\begin{equation}
\langle n_{i\gamma\sigma} \rangle = \frac{n_\gamma}{2}+\sigma\frac{m_\gamma}{2}\cos\left( 
{\bf Q}_1\cdot{\bf r}_i + \phi\right),
\label{ek}
\end{equation}
%
where  $\phi$ is a general phase angle, and  
$n_\gamma$ is the total (spin-up + spin-down)
number of electrons in orbital $\gamma$ at lattice site $i$. The magnetisation  is assumed to point along the $z$-direction, so that 
$
2\langle S_{i\gamma z} \rangle = m_\gamma\cos\left( 
{\bf Q}_1\cdot{\bf r}_i + \phi\right).
$
Our starting point for finding the mean field solution for the magnetic phase is the Hamiltonian,
$
H=H_{TB}+H_{int},
$
where $H_{TB}$ is the kinetic energy (tight-binding) part and $H_{int}$ is the interaction. The latter is given by 
%
\begin{eqnarray}
\bl
H_{int} =
& U \sum_{i\gamma} n_{i\gamma\uparrow} n_{i\gamma\downarrow}
+U'
\sum_{i\sigma{\tilde \sigma} \gamma>\beta} n_{i\gamma\sigma} 
n_{i\beta{\tilde \sigma}}   
\\
&- 2J \sum_{i\gamma>\beta} {\bf S}_{i\gamma}\cdot {\bf S}_{i\beta},
\el
\label{tri}
\end{eqnarray}
%
%
where $\gamma$ is an orbital index and $\sigma$ refers to the spin. Here $U$  and $U'$ are the intra- and inter-orbital Coulomb repulsions, respectively, and $J$ is the Hund's coupling.
Using  Eq. (\ref{ek}), the total mean field Hamiltonian reduces to 
%
\begin{equation}
H^{MF}_{int} = H_U + H_{U'} + H_J,
\label{chatur1}
\end{equation}
%
where 
%
\begin{eqnarray}
\bl
&H_U = -\frac{U}{6}\sum_{\gamma\sigma}\sigma m_\gamma (n_{{\bf Q}\gamma\sigma}e^{i\phi}+n_{-{\bf Q}\gamma\sigma}e^{-i\phi}),  
\\
&H_{U'} = \left( U'-\frac{J}{2} \right)\sum_{{\bf k}\sigma,\gamma\neq\beta}n_\beta a^\dagger_{{\bf k}\gamma\sigma}a_{{\bf k}\gamma\sigma},  
\\
&
H_J = -\frac{J}{4}\sum_{\sigma,\gamma\neq\beta}\sigma m_\beta (n_{{\bf Q}\gamma\sigma}e^{i\phi} + n_{-{\bf Q}\gamma\sigma}e^{-i\phi}),
\label{panch}
\el
\end{eqnarray}
%
and $n_{{\bf Q}\gamma\sigma} = \sum_{\bf k} a^\dagger_{{\bf k}\gamma\sigma}a_{{\bf k}+{\bf Q},\gamma\sigma}$. The magnetic Brillouin zone (MBZ) for  the SDW modulated by the ${\bf Q}_1$- wave vector is shown by the red line in Fig.~\ref{fbzmbzc2c4}(d). \\

The procedure for solving the mean-field Hamiltonian self consistently has been described in~\cite{PhysRevB.79.104510} and consists of the following steps. Initial trial values of $n_\gamma$ and $m_\gamma$ are chosen and the Hamiltonian is diagonalized for all points in the MBZ for these values. The Fermi energy corresponding to the given electronic filling is then determined. Using the eigenvectors, eigenvalues, and the Fermi energy so obtained, new values of   $n_\gamma$ and $m_\gamma$ are calculated. 
These are fed as input into the next iteration and the process is repeated until self-consistency is reached for a given temperature. \\

\begin{figure}
\begin{center}
\includegraphics[width=\linewidth]{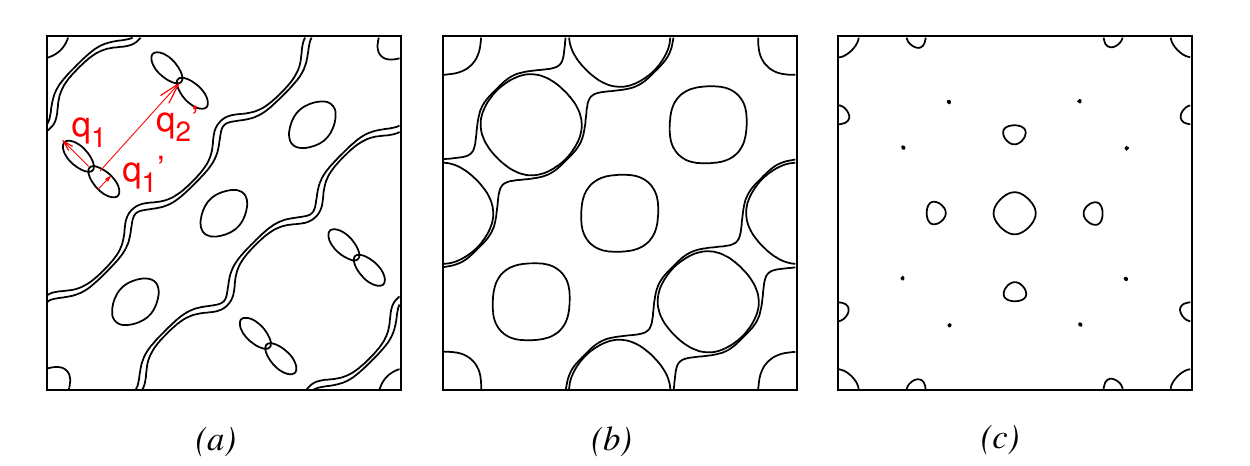}
\end{center}
\vspace{-0.5cm}
\caption{(a) The Fermi surface for $n=6.64$ electrons and $U=2.125$ eV in the full Brillouin zone for the double stripe ($\phi=\pi/4$) magnetic order, showing the electron pocket at the $\Gamma$ point. There are additional electron pockets identified by the wave vectors ${\bf q}_1$, ${\bf q}_1'$ and ${\bf q}_2'$, which can be identified with the scattering wave vectors in the QPI patterns of Fig.~\ref{perkins_qpidisp}. (b) Fermi surfaces for the extended-stripe phase ($\phi=0$) case, and (c) for the orthogonal double stripe (ODS) case.
}
\label{fig2}
\end{figure}

%
\begin{figure}
\centering
\includegraphics[width=1\linewidth, angle=0]{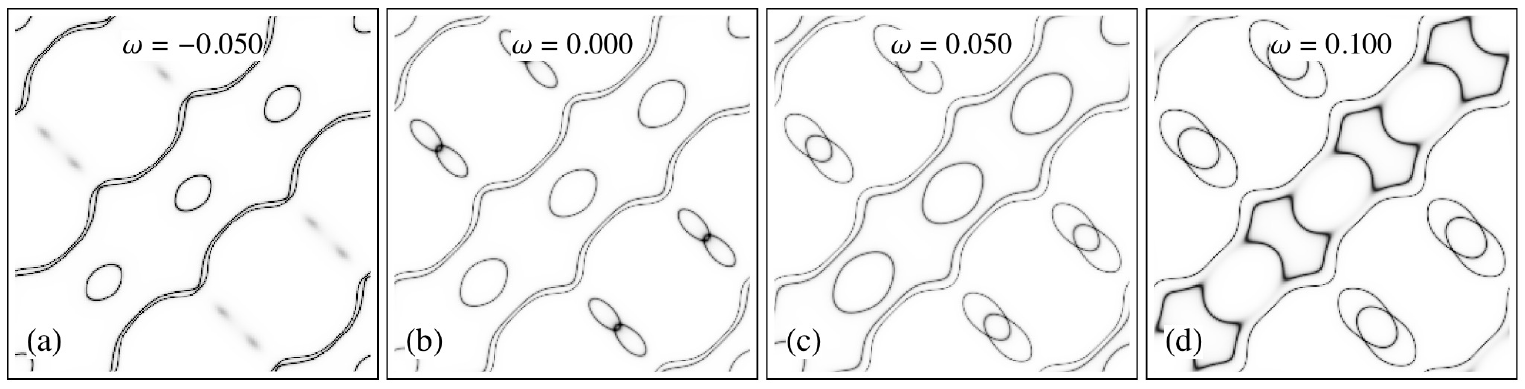}
\includegraphics[width=1\linewidth, angle=0]{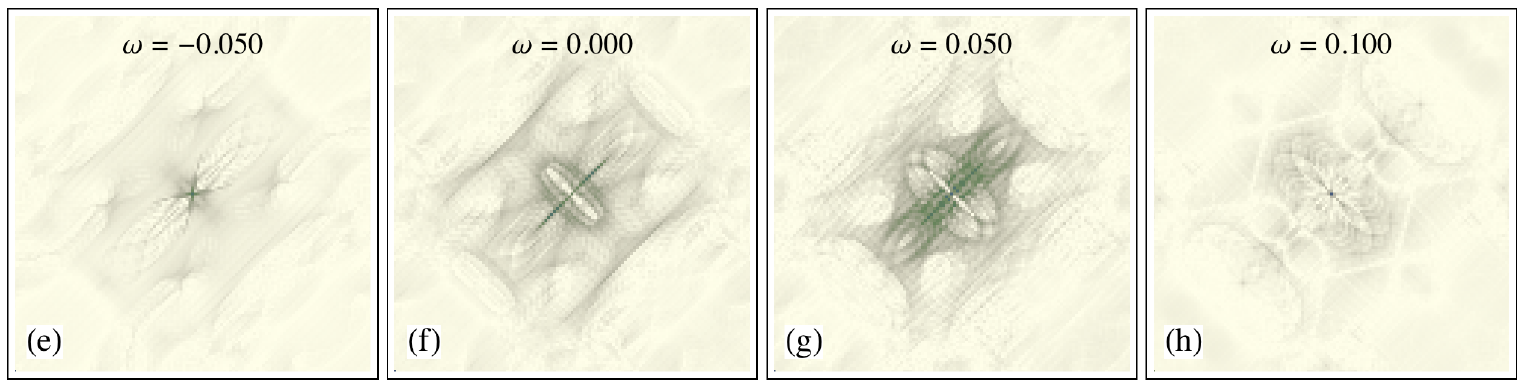}
  \caption{
  (a-d) Spectral intensity maps for energies  -50, 0, 50 and 100meV for the double stripe magnetic order. 
  The electron pockets associated with ${\bf q}_1$, ${\bf q}_1'$ and ${\bf q}_2'$ (shown in Figs.~\ref{fig2} and~\ref{perkins_qpidisp}) 
  appear around -50 meV and increase in size as the energy increases. These wave vectors correspond to the electron-like dispersions in Fig.~\ref{perkins_qpidisp}.
  (e-h) Corresponding QPI maps for energies  -50, 0, 50 and 100meV for the double stripe magnetic order.}
  \label{qpiakwds}
\end{figure}
\begin{figure}
\begin{center}
\includegraphics[width=0.9\linewidth]{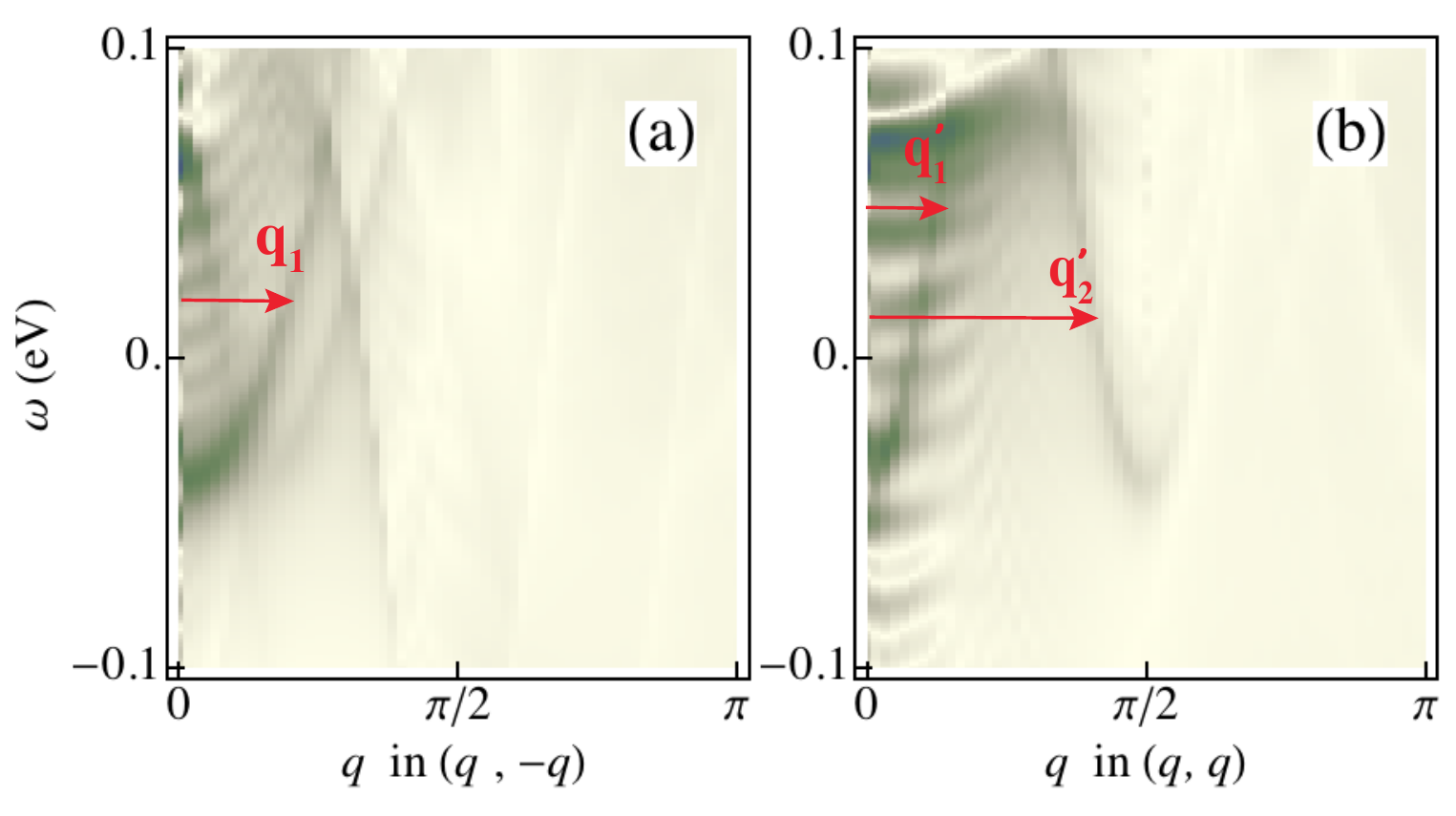}
\end{center}
\vspace*{-0.6cm}
\caption{QPI dispersion along the $(0,0)\rightarrow (-\pi,\pi)$, and $(0,0)\rightarrow (\pi,\pi)$ directions for the double stripe magnetic order. The dominant scattering wave vectors ${\bf q}_1$, ${\bf q}_1'$ and ${\bf q}_2'$ are shown and correspond to the wave vectors connecting the Fermi surfaces in Fig.~\ref{fig2}. 
}
\label{perkins_qpidisp}
\end{figure}

%
\begin{figure}
  \centering
\includegraphics[width=1\linewidth, angle=0]{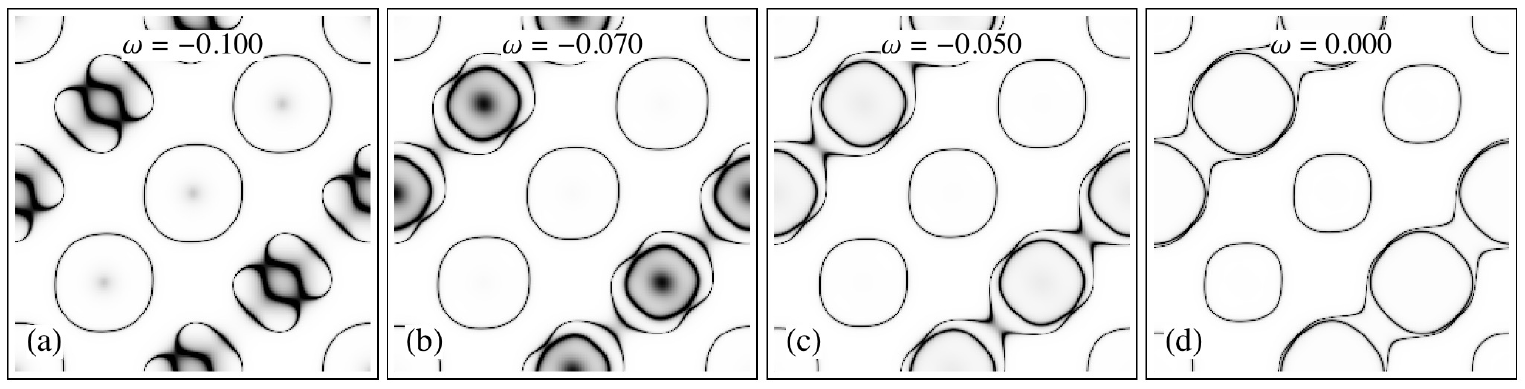}
\includegraphics[width=1\linewidth, angle=0]{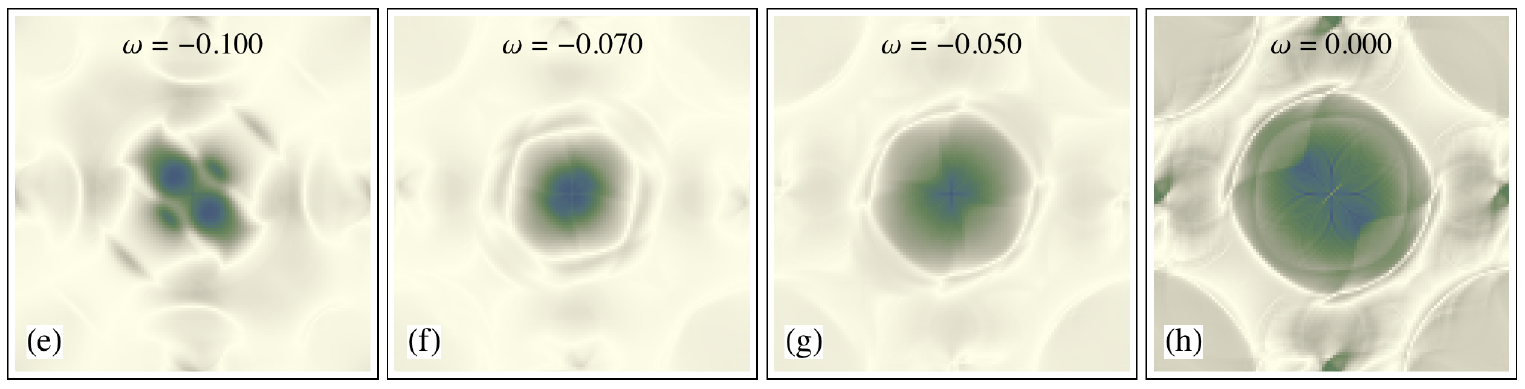}
  \caption{
  (a-d) Spectral intensity maps for four different energies in the full BZ for the extended-stripe ($\phi=0$) case. The electron pocket with the flat cusp is reflected by the sudden appearance of the electron pocket  at -70meV.
  (e-h) QPI maps in the full BZ for four different energies for the extended-stripe case.}
  \label{fig5}
\end{figure}
\begin{figure}
\begin{center}
\includegraphics[width=0.9\linewidth,angle=0]{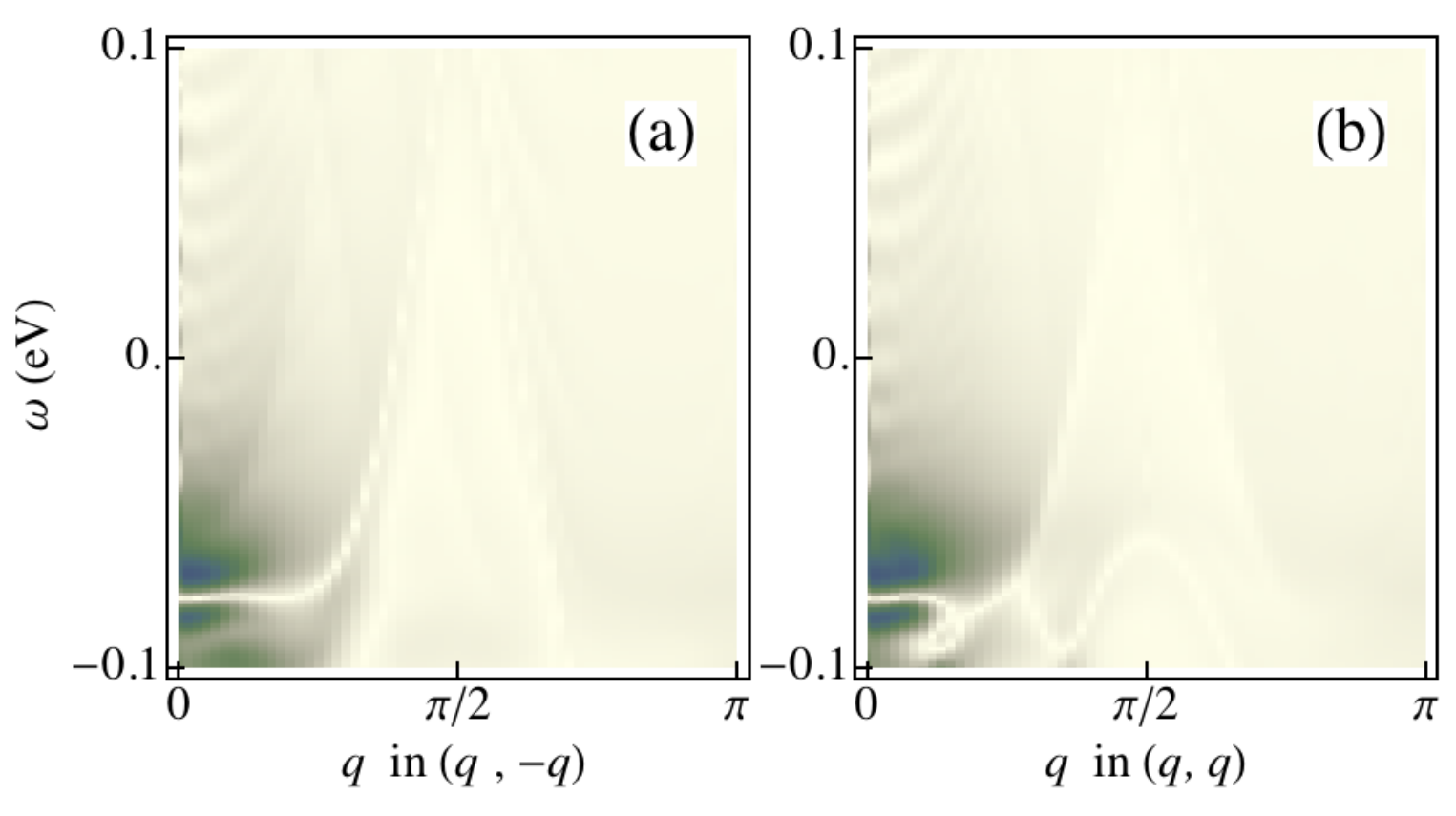}
\end{center}
\vspace*{-0.6cm}
\caption{The QPI dispersion along the (a) $(-\pi,\pi)$ direction and (b) along the $(\pi,\pi)$ direction for $\phi=0$.
}
\label{qpidisp_phi0}
\end{figure}

%
\begin{figure}[h]
  \centering
\includegraphics[width=1\linewidth, angle=0]{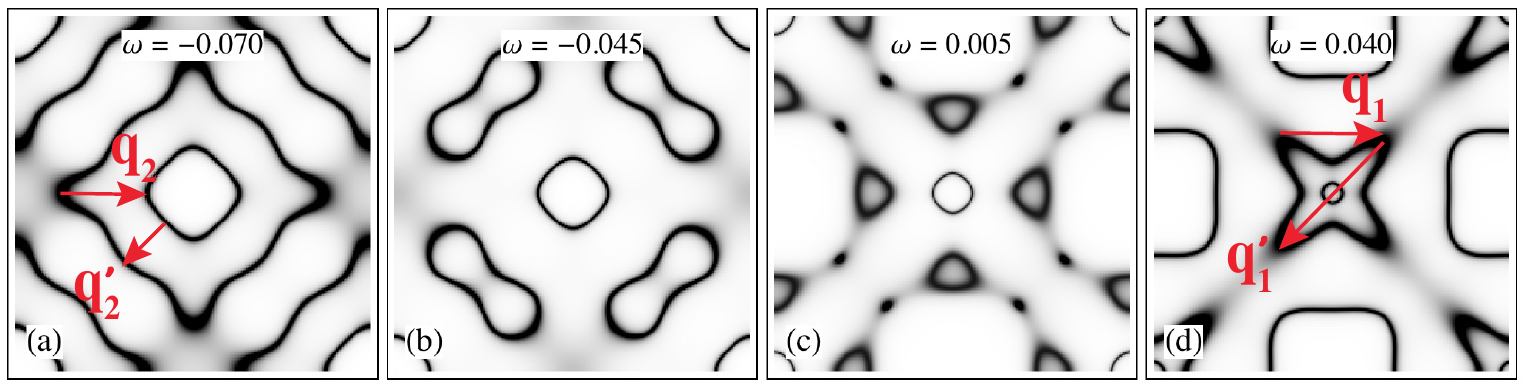}
\includegraphics[width=1\linewidth, angle=0]{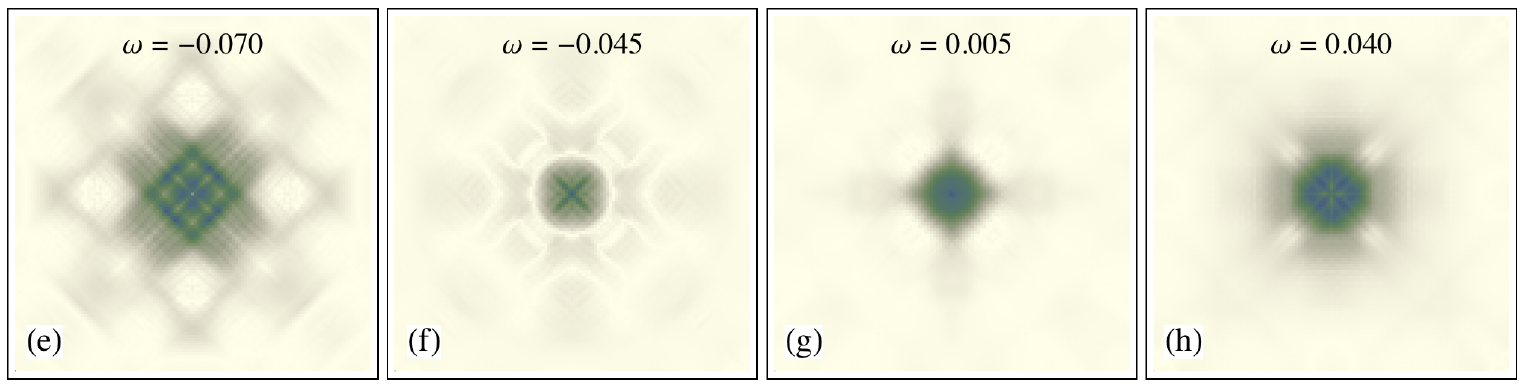}
  \caption{
  (a-d) Spectral intensities at the energies indicated in the BZ double the size of the MBZ.
  (e-h) Corresponding QPI maps for four different energies in the ODS phase in the full BZ.}
  \label{ods_akw}
  \label{ods_qpi}
\end{figure}
\begin{figure}[h]
\begin{center}
\includegraphics[width=0.9\linewidth,angle=0]{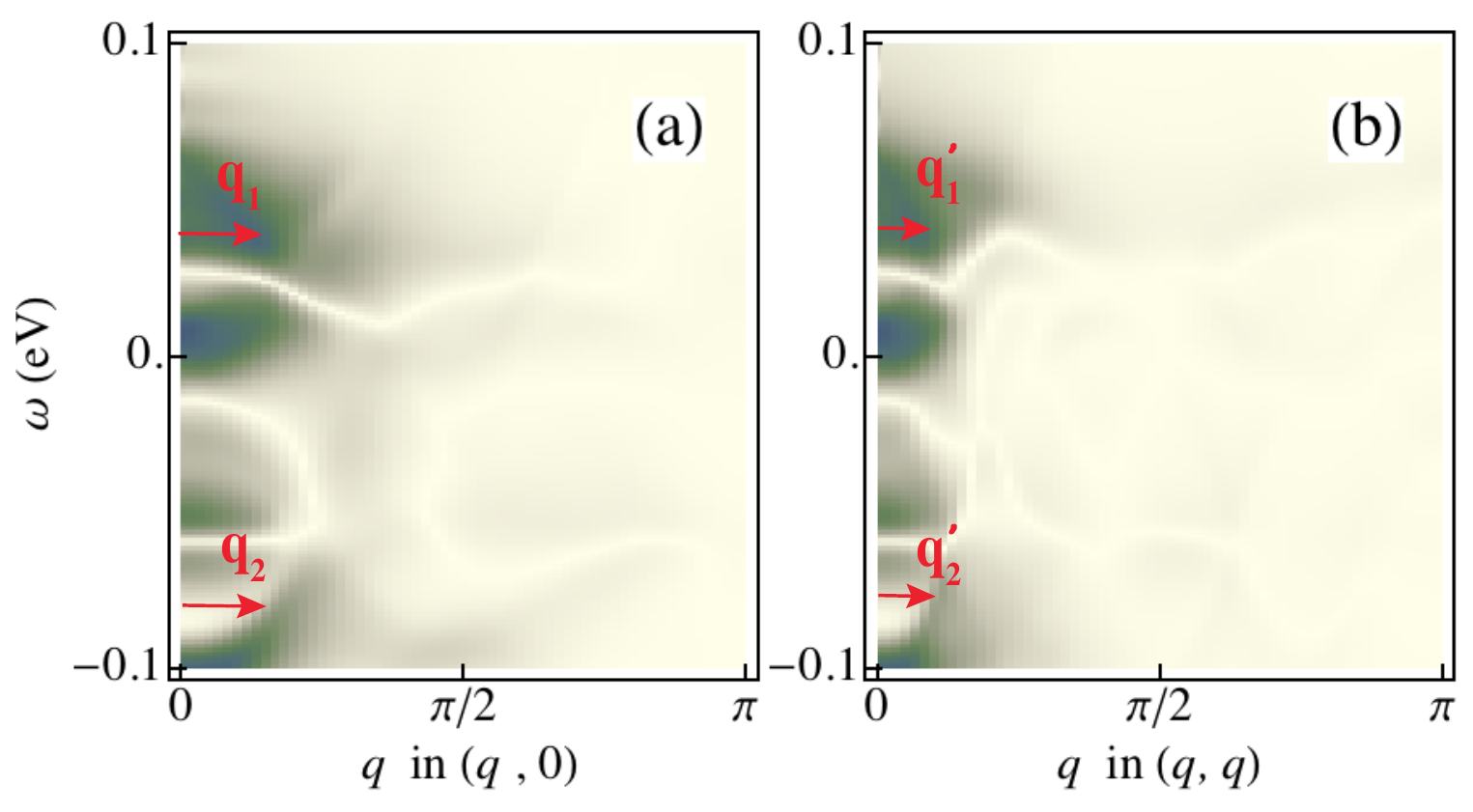}
\end{center}
\vspace*{-0.6cm}
\caption{QPI dispersion along the (a) $(\pi,0)$ direction and (b) $(\pi,\pi)$ direction for the ODS phase. For (a), there is a hole feature (${\bf q}_1$) starting from about 70 meV, an electron feature starting from about 0 meV, a feature at -60 meV, and an electron feature at -100 meV (${\bf q}_2$). There is also a faint hole feature at about -25 meV. For (b), there is a hole feature at 70 meV (${\bf q}_1'$), an electron feature at 0 meV, a feature at -60 meV   and an electron feature at -100 meV (${\bf q}_2'$). 
}
\label{ods_qpidisp_both}
\end{figure}

{\it Theoretical formulation of QPI}:
The scattering off a magnetic impurity is modelled by the Hamiltonian
%
\begin{equation}
H_{imp}=\sum_{{\bf k}{\bf k'}\mu\nu\sigma\sigma'}c^{\dagger}_{\mu{\bf k}\sigma}(J^{\mu\nu}_{\sigma\sigma'}{\bf S}\cdot {\mbox{{\boldmath$\sigma$}}}_{\sigma\sigma'})c_{\nu{\bf k'}\sigma'},
\end{equation}
%
where ${\bf S}$ is the impurity spin. We take impurity spin along $z$-direction and assume purely localized impurity scattering. 
Introducing the matrices 
${\hat U} = \sigma_z \otimes {\hat J}$, 
 and 
${\hat T}
=[1-{\hat U}\sum_{\bf k}G_0({\bf k},\omega)]^{-1}{\hat U}
$,
 the QPI under the $T$-matrix approximation is given by~\cite{Balatsky:2006aa}
%
\begin{eqnarray}
\bl %
 {\rm QPI} 
 \sim
 {\rm Im}
 \sum_{{\bf k}\sigma}
  {\rm Tr} 
  \Big[
  G_{0\sigma}({\bf k},\omega){\hat T}_\sigma G_{0\sigma}({\bf k}+{\bf q},\omega)
  \Big],
  \el
\end{eqnarray}
%
where  $\otimes$ is the direct product and ${\hat J}$ is a matrix in the purely orbital basis: ${\hat J}_{\mu\nu}=J^{\mu\nu}_{\uparrow\uparrow}=J^{\mu\nu}_{\downarrow\downarrow}$.\\

{\it Numerical Results for the double stripe phase:} 
This particular magnetic structure is exhibited by the iron-chalcogenides and is obtained by putting $\phi=\pi/4$ in Eq.~(\ref{ek}). In this section we present numerical results for Fe$_{1+y}$Te in the mean-field formulation. We employ the 5-orbital tight binding model of~\cite{PhysRevB.90.165123} and consider Fe$_{1.08}$Te which, assuming that each excess Fe atom contributes eight electrons corresponds to an electronic density of $n=6.64$. 
We should emphasise that the  chemical potential shifts by about 0.4 eV in the $y=0.08$ case  compared to $y=0$
and that the hole pocket at $\Gamma$ is replaced by an electron pocket~\cite{PhysRevB.90.165123}. 
 This latter feature has been observed by ARPES in other electron doped Fe compounds as well~\cite{PhysRevB.84.020509}.

For the mean field calculation we fix $J=0.25U$ and carry out the self-consistent calculation for the  magnetization.
  The magnetic moment in Fe$_{1+y}$Te is around $2.1\mu_B$~\cite{PhysRevB.79.054503}. Further, from LDA and ARPES, the band structure in the magnetic phase of Fe$_{1+y}$Te is known to have an electron pocket at the $\Gamma$ point~\cite{PhysRevLett.111.217002}. From our mean-field analysis, a magnetic moment of $2.1 \mu_B$ corresponds to about $2.2$~eV. There is a hole like feature at the $\Gamma$ point for this value of $U$, which is replaced by an electron like feature at lower values. 
  Thus in order to maintain consistency of our results with  LDA and ARPES~\cite{PhysRevLett.111.217002}, we choose $U=2.125$~eV. The magentization has a somewhat lower value of  $1.02\mu_B$ for this interaction, but the Fermi surface so obtained indeed shows an electron pocket at $\Gamma$ (see Fig.~\ref{fig2}(a)).

The spectral functions for four different energies and the corresponding QPI maps are shown in Fig.~\ref{qpiakwds}. 
One can trace the main momentum scattering vectors, shown by the  ${\bf q}_i$ in Fig.~\ref{fig2}(a), in the QPI maps  of Fig.~\ref{qpiakwds}.

In Fig.~\ref{perkins_qpidisp} we show the ``QPI dispersion'',  i.e. the QPI intensities as a function of frequency and momentum,  along the $(\pi,-\pi)$- and  $(\pi,\pi)$-directions in the full Brillouin zone (BZ). 
In Fig.~\ref{perkins_qpidisp} we see an electron like feature starting from about -0.05 eV along the $(-\pi,\pi)$ direction. We have labeled the main momentum characterizing this by ${\bf q}_1$. We also have an electron like feature starting from the same energy, but at lower momenta, ${\bf q}_1'$, along the $(\pi,\pi)$ direction. Along the latter direction there is also an electron-like feature starting from the same energy but at higher momenta, ${\bf q}_2'$. The origin of these features is clear from Fig.~\ref{fig2} (a) where we have labeled these momenta on the Fermi surface. Thus these features can be related to the ``twin'' electron pockets connected by ${\bf q}_2'$. These ``twin'' pockets indeed start at an energy about -0.05 eV as can be seen from Fig.~\ref{qpiakwds}.\\

{\it Numerical Results for the extended-stripe phase:}
The extended-stripe phase  is obtained by setting $\phi=0$ in Eqs.~(\ref{ek}-\ref{panch}). The corresponding magnetic structure is shown in Fig.~\ref{fbzmbzc2c4}(b) and consists of two interpenetrating antiferromagnetic sublattices and a sublattice with zero magnetic moment. We repeat the self-consistent calculation for $n=6.64$ electrons. To have a magnetization similar to the $\phi=\pi/4$ case, we choose in this case $U=2$ eV, for which the magnetization equals  $0.91\mu_B$. The corresponding Fermi surface is shown in Fig.~\ref{fig2}(b). The Fermi surface consists of a hole pocket at the $\Gamma$ point, and electron-pockets with a relatively flat cusp at  ${\bf Q}_2$. 

Fig.~\ref{fig5} shows the spectral intensities at four different energies in the full BZ. The electron pocket with the flat cusp makes a sudden appearance  at around -70 meV around the $(-\pi/2,\pi/2)$ point (see Fig.~\ref{fig5}(b)). Furthermore Fig.~\ref{fig5} shows the corresponding QPI maps for these energies in the full BZ. At -70 meV there is an abrupt qualitative shift in the QPI map around $\Gamma$, coincident with the appearance of the electron pocket with the flat cusp. In Fig.~\ref{qpidisp_phi0} we have shown the QPI dispersion along the $(\pi,\pi)$ and $(-\pi,\pi)$ directions. Along both directions we see an electron like feature with a heavy mass which corresponds to the electron pocket with the broad cusp in the quasiparticle dispersion. 
This flat pocket is also present in the DS phase, however there it is located at a much lower energy than that considered in the QPI dispersion here and hence does not show up for that case.

{\it The orthogonal double stripe phase:}
This phase is shown in Fig.~\ref{fbzmbzc2c4}(c) and has been discussed in Ref.~\cite{PhysRevLett.109.157206}. The MBZ is the small diamond in the center in Fig.~\ref{fbzmbzc2c4}(e). Unlike the other two phases which had $C_2$ symmetry  this phase possesses $C_4$-tetragonal symmetry. It can be described by the following mean-field ansatz:
\begin{equation}
\langle n_{i\gamma\sigma} \rangle = \frac{n_\gamma}{2}+\sigma\frac{m_\gamma}{2}
[
 \sin\left( 
{\bf Q}_1\cdot{\bf r}_i\right)  + \cos  \left( 
{\bf Q}_2\cdot{\bf r}_i\right)
],
\label{ek_ods}
\end{equation}
%
In this case the magnetization at site $i$ is given by
$
\langle S_{i\gamma z}  \rangle =  m_\gamma \left(  \sin {\bf Q}_1\cdot{\bf r}_i + \cos {\bf Q}_2\cdot{\bf r}_i\right)/2
$.
%
Under this ansatz the mean field Hamiltonian for the interaction is given by Eq.~(\ref{chatur1}) with
%
\begin{eqnarray}
\bl
H_U 
&= 
\frac{iU}{6}
\sum_{\sigma  \gamma}
\sigma m_{\gamma} 
(n_{{\bf Q}_1\gamma\sigma}-n_{{\bf Q}_2\gamma\sigma}) +h.c.,
 \\
H_J & = \frac{J}{4}
\sum_{\sigma,\gamma\neq\beta}
\sigma m_\beta(i n_{{\bf Q}_1\gamma\sigma}-n_{{\bf Q}_2\gamma\sigma})+h.c.,
\el
\end{eqnarray}
%
and $H_{U'}$ being the same as earlier.

Based on the self-consistent results for the magnetization as a function of $U$, with $n=6$ and $J=0.25U$, we choose $U=2$ eV which corresponds to a magnetization of 2.54$\mu_B$. The Fermi surfaces are shown in Fig.~\ref{fig2}(c) in a region twice the size of the MBZ and rotated by 45$^{\rm o}$. The Fermi surface consists of a hole pocket at $\Gamma$. We show the spectral intensities at four different energies in Fig.~\ref{ods_akw}. The corresponding QPI maps for these energies are also shown in Fig.~\ref{ods_qpi}, while the QPI dispersion is shown in Fig.~\ref{ods_qpidisp_both}. Figure~\ref{ods_qpidisp_both}(a) shows the cut along the $(\pi,0)$ direction and Fig.~\ref{ods_qpidisp_both}(b) along the $(\pi,\pi)$ direction.

There are several features in the spectral intensities which give rise to a more complex QPI pattern and QPI dispersion compared to the DS and extended-stripe cases.  
 In Fig.~\ref{ods_qpidisp_both}(a) there is a hole feature starting from about 70meV, an electron feature starting from about 0meV, a feature at -60meV, and an electron feature at -100meV. There is also a faint hole feature at about -25meV. All of these can be understood in terms of the spectral intensities of Fig.~\ref{ods_akw}. 
 As an example, the hole dispersion marked by ${\bf q}_1$ and the electron dispersion marked by ${\bf q}_2$ in Fig.~\ref{ods_qpidisp_both}(a) can be associated with ${\bf q}_1$ and ${\bf q}_2$ in Fig.~\ref{ods_akw}(d) and (a) respectively. Similarly ${\bf q}_1'$ and ${\bf q}_2'$ in Fig.~\ref{ods_qpidisp_both}(b) can be associated with ${\bf q}_1'$ and ${\bf q}_2'$ in Fig.~\ref{ods_akw}(d) and (a) respectively.

In this article, using QPI as a possible tool to detect the magnetic structure based on scattering of quasipartciles from nonmagnetic impurities, we theoretically investigated three different possible magnetic phases for the parent compounds of the iron-chalcogenide superconductors. 
Most importantly, we have shown that different magnetic phases realized in iron chalocgenides can be easily identified by means of QPI from non-magnetic impurities due to very different electronic structure reconstruction in all these phases.
The first phase considered was the experimentally observed DS phase, which has a SDW wave vector ${\bf Q}_1$ and a phase angle of $\phi=\pi/4$.   The second phase  was the extended stripe phase which has $\phi=0$, and the third phase was the ODS phase which consists of both ${\bf Q}_1$ and ${\bf Q}_2$ wave vectors. We used a tight binding model developed for Fe$_{1+y}$Te and a multi-orbital interaction term to obtain the magnetic states self-consistently. For the first two states we considered an electronic filling of 6.64 electrons and a magnetic moment of about 1$\mu_B$ to match the electron pocket experimentally seen for the DS phase. For the ODS phase we chose 6 electrons and a magnetic moment of 2.5$\mu_B$. The QPI calculations were carried out using the $T$-matrix approximation.

In the DS phase, the Fermi surfaces consist of an electron pocket at $\Gamma$ and twin-electron pockets at the corners of the magnetic BZ. The QPI features are mainly dominated by these twin pockets as seen in the QPI dispersion plots which predominantly show an electron-type dispersion. For the extended stripe phase, the Fermi surface consists of a hole pocket at $\Gamma$, and an electron type dispersion with a flat cusp in the $(-\pi,\pi)$ direction. It is predominantly this feature which is reflected in the QPI behavior. The scattering across this electron-pocket dominates the QPI intensity map and results in the electron-feature in the QPI dispersion.

The ODS phase was relatively more complicated than the other two phases considered. The scatterings between various contours of constant energy resulted in the features seen in the QPI dispersion.\\

{\it Acknowledgments:}
We thank  P. Wahl,  A. Yaresko, P. Thalmeier, M.N. Gastiasoro,  B.M. Andersen, and M. Enayat for helpful comments and  for useful discussions. We are grateful to the Max Planck Institute for the Physics of Complex Systems (MPI-PKS) for the use of computer facilities.
B.K. and  A.A. wish to acknowledge the Korea Ministry of Education, Science and Technology, Gyeongsangbuk-Do and Pohang City for Independent Junior Research Groups at the Asia Pacific Center for Theoretical Physics. 
The work by B.K. and  A.A. was supported through  NRF funded by MSIP of Korea (2015R1C1A1A01052411).  A.A. acknowledges support by  Max Planck POSTECH / KOREA Research Initiative (No. 2011-0031558) programs through NRF funded by MSIP of Korea. 
The work of IE was supported by the Focus Program 1458
Eisen-Pniktide of the DFG. IE acknowledges support by the Ministry of
Education and Science of the Russian Federation in the
framework of Increase Competitiveness Program of NUST
MISiS (N 2-2014-015).

\bibliography{./calc}

\begin{thebibliography}{10}

\bibitem{RevModPhys.83.1589}
G.~R. Stewart, 2011 \emph{Rev. Mod. Phys.}, 83 1589--1652.

\bibitem{PhysRevLett.100.237003}
D.~J. Singh and M.-H. Du, 2008 \emph{Phys. Rev. Lett.}, 100 237003.

\bibitem{PhysRevLett.101.026403}
L.~Boeri, O.~V. Dolgov, and A.~A. Golubov, 2008 \emph{Phys. Rev. Lett.}, 101
  026403.

\bibitem{PhysRevLett.101.057003}
I.~I. Mazin, D.~J. Singh, M.~D. Johannes, and M.~H. Du, 2008 \emph{Phys. Rev.
  Lett.}, 101 057003.

\bibitem{PhysRevLett.101.177005}
C.~Liu, G.~D. Samolyuk, Y.~Lee, N.~Ni, T.~Kondo, A.~F. Santander-Syro, S.~L.
  Bud'ko, J.~L. McChesney, E.~Rotenberg, T.~Valla, A.~V. Fedorov, P.~C.
  Canfield, B.~N. Harmon, and A.~Kaminski, 2008 \emph{Phys. Rev. Lett.}, 101
  177005.

\bibitem{PhysRevB.79.054517}
D.~V. Evtushinsky, D.~S. Inosov, V.~B. Zabolotnyy, A.~Koitzsch, M.~Knupfer,
  B.~B\"uchner, M.~S. Viazovska, G.~L. Sun, V.~Hinkov, A.~V. Boris, C.~T. Lin,
  B.~Keimer, A.~Varykhalov, A.~A. Kordyuk, and S.~V. Borisenko, 2009
  \emph{Phys. Rev. B}, 79 054517.

\bibitem{0295-5075-83-4-47001}
H.~Ding, P.~Richard, K.~Nakayama, K.~Sugawara, T.~Arakane, Y.~Sekiba,
  A.~Takayama, S.~Souma, T.~Sato, T.~Takahashi, Z.~Wang, X.~Dai, Z.~Fang, G.~F.
  Chen, J.~L. Luo, and N.~L. Wang, 2008 \emph{EPL (Europhysics Letters)}, 83(4)
  47001.

\bibitem{PhysRevLett.101.216402}
A.~I. Coldea, J.~D. Fletcher, A.~Carrington, J.~G. Analytis, A.~F. Bangura,
  J.-H. Chu, A.~S. Erickson, I.~R. Fisher, N.~E. Hussey, and R.~D. McDonald,
  2008 \emph{Phys. Rev. Lett.}, 101 216402.

\bibitem{PhysRevB.81.024511}
I.~Eremin and A.~V. Chubukov, 2010 \emph{Phys. Rev. B}, 81 024511.

\bibitem{PhysRevB.79.054503}
S.~Li, C.~de~la Cruz, Q.~Huang, Y.~Chen, J.~W. Lynn, J.~Hu, Y.-L. Huang, F.-C.
  Hsu, K.-W. Yeh, M.-K. Wu, and P.~Dai, 2009 \emph{Phys. Rev. B}, 79 054503.

\bibitem{PhysRevLett.109.157206}
S.~Ducatman, N.~B. Perkins, and A.~Chubukov, 2012 \emph{Phys. Rev. Lett.}, 109
  157206.

\bibitem{PhysRevLett.107.216403}
I.~A. Zaliznyak, Z.~Xu, J.~M. Tranquada, G.~Gu, A.~M. Tsvelik, and M.~B. Stone,
  2011 \emph{Phys. Rev. Lett.}, 107 216403.

\bibitem{1742-6596-200-2-022058}
P.~Sindzingre, N.~Shannon, and T.~Momoi, 2010 \emph{Journal of Physics:
  Conference Series}, 200(2) 022058.

\bibitem{PhysRevB.86.134512}
Y.-Y. Tai, J.-X. Zhu, M.~J. Graf, and C.~S. Ting, 2012 \emph{Phys. Rev. B}, 86
  134512.

\bibitem{Enayat08082014}
M.~Enayat, Z.~Sun, U.~R. Singh, R.~Aluru, S.~Schmaus, A.~Yaresko, Y.~Liu,
  C.~Lin, V.~Tsurkan, A.~Loidl, J.~Deisenhofer, and P.~Wahl, 2014
  \emph{Science}, 345(6197) 653--656.

\bibitem{PhysRevLett.102.247001}
W.~Bao, Y.~Qiu, Q.~Huang, M.~A. Green, P.~Zajdel, M.~R. Fitzsimmons,
  M.~Zhernenkov, S.~Chang, M.~Fang, B.~Qian, E.~K. Vehstedt, J.~Yang, H.~M.
  Pham, L.~Spinu, and Z.~Q. Mao, 2009 \emph{Phys. Rev. Lett.}, 102 247001.

\bibitem{PhysRevLett.115.177203}
P.~Materne, C.~Koz, U.~K. R\"o\ss{}ler, M.~Doerr, T.~Goltz, H.~H. Klauss,
  U.~Schwarz, S.~Wirth, and S.~R\"o\ss{}ler, 2015 \emph{Phys. Rev. Lett.}, 115
  177203.

\bibitem{0034-4885-74-12-124508}
P.~J. Hirschfeld, M.~M. Korshunov, and I.~I. Mazin, 2011 \emph{Reports on
  Progress in Physics}, 74(12) 124508.

\bibitem{0034-4885-74-12-124513}
J.~E. Hoffman, 2011 \emph{Reports on Progress in Physics}, 74(12) 124513.

\bibitem{PhysRevLett.71.3363}
J.~M. Byers, M.~E. Flatt\'e, and D.~J. Scalapino, 1993 \emph{Phys. Rev. Lett.},
  71 3363--3366.

\bibitem{0295-5075-85-3-37005}
F.~Wang, H.~Zhai, and D.-H. Lee, 2009 \emph{EPL (Europhysics Letters)}, 85(3)
  37005.

\bibitem{Hanaguri23042010}
T.~Hanaguri, S.~Niitaka, K.~Kuroki, and H.~Takagi, 2010 \emph{Science},
  328(5977) 474--476.

\bibitem{Hanaguri13022009}
T.~Hanaguri, Y.~Kohsaka, M.~Ono, M.~Maltseva, P.~Coleman, I.~Yamada, M.~Azuma,
  M.~Takano, K.~Ohishi, and H.~Takagi, 2009 \emph{Science}, 323(5916) 923--926.

\bibitem{Allan:2012aa}
M.~P. Allan, A.~W. Rost, A.~P. Mackenzie, Y.~Xie, J.~C. Davis, K.~Kihou, C.~H.
  Lee, A.~Iyo, H.~Eisaki, and T.-M. Chuang, 2012 \emph{Science}, 336(6081)
  563--567.

\bibitem{Chi:2014aa}
S.~Chi, S.~Johnston, G.~Levy, S.~Grothe, R.~Szedlak, B.~Ludbrook, R.~Liang,
  P.~Dosanjh, S.~A. Burke, A.~Damascelli, D.~A. Bonn, W.~N. Hardy, and
  Y.~Pennec, 2014 \emph{Phys. Rev. B}, 89 104522.

\bibitem{Akbari:2010aa}
A.~Akbari, J.~Knolle, I.~Eremin, and R.~Moessner, 2010 \emph{Phys. Rev. B}, 82
  224506.

\bibitem{PhysRevLett.104.257001}
J.~Knolle, I.~Eremin, A.~Akbari, and R.~Moessner, 2010 \emph{Phys. Rev. Lett.},
  104 257001.

\bibitem{Yamakawa:2015aa}
Y.~Yamakawa and H.~Kontani, 2015 \emph{Phys. Rev. B}, 92 045124.

\bibitem{Hirschfeld:2015aa}
P.~J. Hirschfeld, D.~Altenfeld, I.~Eremin, and I.~I. Mazin, 2015 \emph{Phys.
  Rev. B}, 92 184513.

\bibitem{Sykora:2011aa}
S.~Sykora and P.~Coleman, 2011 \emph{Phys. Rev. B}, 84 054501.

\bibitem{Akbari:2011aa}
A.~Akbari, P.~Thalmeier, and I.~Eremin, 2011 \emph{Phys. Rev. B}, 84 134505.

\bibitem{0295-5075-100-3-37004}
J.-X. Zhu and A.~R. Bishop, 2012 \emph{EPL (Europhysics Letters)}, 100(3)
  37004.

\bibitem{Akbari:2013aa}
A.~Akbari and P.~Thalmeier, 2013 \emph{EPL (Europhysics Letters)}, 102(5)
  57008.

\bibitem{0295-5075-103-2-27004}
J.~Li, Y.-H. Chen, and C.~S. Ting, 2013 \emph{EPL (Europhysics Letters)},
  103(2) 27004.

\bibitem{0295-5075-100-3-37002}
Y.~Gao, H.~X. Huang, and P.~Q. Tong, 2012 \emph{EPL (Europhysics Letters)},
  100(3) 37002.

\bibitem{Akbari:2014aa}
A.~Akbari and P.~Thalmeier, 2014 \emph{EPL (Europhysics Letters)}, 106(2)
  27006.

\bibitem{doi:10.7566/JPSJ.83.061015}
I.~Eremin, J.~Knolle, R.~M. Fernandes, J.~Schmalian, and A.~V. Chubukov, 2014
  \emph{Journal of the Physical Society of Japan}, 83(6) 061015.

\bibitem{McElroy1}
K.~McElroy, R.~W. Simmonds, J.~E. Hoffman, D.~H. Lee, J.~Orenstein, H.~Eisaki,
  S.~Uchida, and J.~C. Davis, 2003 \emph{Nature}, 422(6932) 592--596.

\bibitem{PhysRevB.90.165123}
S.~Ducatman, R.~M. Fernandes, and N.~B. Perkins, 2014 \emph{Phys. Rev. B}, 90
  165123.

\bibitem{PhysRevB.79.104510}
R.~Yu, K.~T. Trinh, A.~Moreo, M.~Daghofer, J.~A. Riera, S.~Haas, and
  E.~Dagotto, 2009 \emph{Phys. Rev. B}, 79 104510.

\bibitem{Balatsky:2006aa}
A.~V. Balatsky, I.~Vekhter, and J.-X. Zhu, 2006 \emph{Rev. Mod. Phys.}, 78
  373--433.

\bibitem{PhysRevB.84.020509}
C.~Liu, A.~D. Palczewski, R.~S. Dhaka, T.~Kondo, R.~M. Fernandes, E.~D. Mun,
  H.~Hodovanets, A.~N. Thaler, J.~Schmalian, S.~L. Bud'ko, P.~C. Canfield, and
  A.~Kaminski, 2011 \emph{Phys. Rev. B}, 84 020509.

\bibitem{PhysRevLett.111.217002}
P.-H. Lin, Y.~Texier, A.~Taleb-Ibrahimi, P.~Le~F\`evre, F.~Bertran,
  E.~Giannini, M.~Grioni, and V.~Brouet, 2013 \emph{Phys. Rev. Lett.}, 111
  217002.

\end{thebibliography}
\bibliographystyle{epl3}
\end{document}